\newcommand\ignore[1]{}
\documentclass[prl,twocolumn,showpacs,preprintnumbers,amsmath,amssymb,superscriptaddress,nofootinbib]{revtex4}

\usepackage{color}
\newcommand{\comment}[1]{}   

\usepackage{cancel}
\usepackage{graphicx}
\usepackage{dcolumn}
\usepackage{bm}

\begin{document}

\title{Hybrid Monte Carlo Simulation of Graphene on the Hexagonal Lattice}

\author{R. C. Brower}
\affiliation{Department of Electrical and Computer Engineering,  
Boston University, 8 St. Mary Street, Boston,  MA 02215,  
United States of America} 
\affiliation{Department of Physics,  Boston University,
 590 Commonwealth Avenue, Boston,  MA 02215,  United States of America}
\affiliation{Center for Computational Science,  Boston University,
 3 Cummington Street, Boston,  MA 02215,  United States of America}
\author{C. Rebbi}
\affiliation{Department of Physics,  Boston University,
 590 Commonwealth Avenue, Boston,  MA 02215,  United States of America}
\affiliation{Center for Computational Science,  Boston University,
 3 Cummington Street, Boston,  MA 02215,  United States of America}
\author{D. Schaich}
\affiliation{Department of Physics,  Boston University,
 590 Commonwealth Avenue, Boston,  MA 02215,  United States of America}
\affiliation{Center for Computational Science,  Boston University,
 3 Cummington Street, Boston,  MA 02215,  United States of America}

\date{\today}

\begin{abstract}
We present a method for direct hybrid Monte Carlo simulation of 
graphene on the hexagonal lattice.  We compare the results of the
simulation with exact results for a unit hexagonal cell system, where the 
Hamiltonian can be solved analytically.
\end{abstract}

\maketitle
\section{Introduction}

Graphene, a single layer of carbon atoms forming a hexagonal lattice, has
remarkable properties~\cite{graphene}.  In the tight-binding approximation
the quadratic Hamiltonian gives origin to a dispersion formula which for
low momenta is analogous to the dispersion formula for relativistic
fermions in two dimensions~\cite{dispersion}.  This has prompted some 
researchers to adapt to graphene lattice gauge theory techniques 
which have been profitably used for the study of Quantum Chromodynamics 
and other particle systems~\cite{LGT}.  In the work of~\cite{LGT-graphene}
one approximates first the
tight-binding Hamiltonian of graphene with a Dirac Hamiltonian,
incorporates the Coulomb interaction through the introduction of a suitable
electromagnetic field, and finally discretizes the resulting continuum
quantum field theory on a hypercubic space-time lattice.  The hybrid
Monte Carlo method~\cite{HMC}, widely used in lattice gauge theory to simulate fermions interacting
with quantum gauge fields, can then be used to investigate
the effects of the Coulomb interaction in the graphene system.

The approach outlined above has led to very interesting and valuable
results~\cite{LGT-graphene}, yet one would think that, since the
starting point is a system already defined on a lattice, it should be
possible to apply the hybrid Monte Carlo technique directly to the
graphene lattice. The clear advantage of this approach is the
  direct connection to the experimentally determined physical lattice
  constants of the tight-binding model, which represents
  an accurate description of the experimental system.  In this letter
we illustrate how this can be done.

\begin{figure}
  \includegraphics[width=5cm]{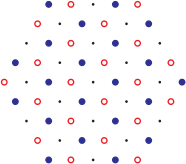}
  \caption{The hexagonal graphene lattice consists of the two triangular sublattices $A$ (solid) and $B$ (empty), which together with the centers of the hexagons (dots) form a finer, underlying triangular lattice.}
  \label{sublattices}
\end{figure}

Graphene is a system of interacting electrons located at the vertices
of a hexagonal lattice.  It is convenient to think of the graphene
lattice as consisting of two triangular sublattices, which we denote
by $A$ and $B$, which together with the centers of the hexagons
(sublattice $C$) form a finer, underlying triangular lattice
(Fig.~\ref{sublattices}).  We introduce fermionic annihilation and
creation operators $a_{x,s}, a^\dag_{x,s}$ for the electrons on the
two sublattices, where $x$ is a site index and $s= \pm 1$
is the spin index. The lattice must be made finite in order to
perform numerical simulations. While there is a broad range of boundary
conditions of physical interest, here we consider periodic systems formed
by identifying opposing sides of a hexagonal lattice of length $L$,
illustrated in Fig.~\ref{sublattices} for $L = 4$.

The tight-binding Hamiltonian $H$ consists of two terms: the quadratic
kinetic term, 
\begin{equation}
H_2=\sum_{\langle x,y\rangle,s} -\kappa (a^\dag_{x,s}a_{y,s} + \mbox{h.c.}),
\label{eq2}
\end{equation}
where the sum runs over all pairs $\langle x,y\rangle$ of
nearest neighbor sites (coupling the $A$ and $B$ sublattices) and the two values of the spin,
and the Coulomb interaction Hamiltonian,
\begin{equation}
H_C=\sum_{x,y} e^2 V_{x,y} q_x q_y,
\label{eq3}
\end{equation}
where 
\begin{equation}
q_x=a^\dag_{x,1} a_{x,1}+a^\dag_{x,-1} a_{x,-1} -1
\label{eq4}
\end{equation}
is a local charge operator and $V$ is the interaction potential.  We have
explicitly introduced the charge coupling constant $e$.

Several comments are in order. First note that in the kinetic term we have neglected
the smaller next-to-nearest neighbor hopping within each
sublattice, which would introduce a small (probably manageable)
complex phase in the path integral. The charge operator Eq.~\ref{eq4} has
a $-1$ to account for the background charge of the
carbon ion: it ensures that the system is neutral at half filling,
and it will play an important role for our functional integral
formulation.  $V$ could be the actual 3d Coulomb potential, but could
be any other interaction potential.  The only thing crucial for us
is that the matrix $V_{x,y}$ be positive definite.  Finally, we note
that the Hamiltonian of Eqs.~\ref{eq2}-\ref{eq3} commutes with the
isospin generators
\begin{align}
  \label{eq:iso}
  I_{\pm} & = a^\dag_{x,s}\sigma^{ss'}_{\pm} a_{x,s'}, &
  I_3 & = a^\dag_{x,s} \sigma^{ss'}_3 a_{x,s'}/2.
\end{align}

In order to explore the properties of the system one would like to calculate
expectation values,
\begin{equation}
\langle {\cal O}_1(t_1) {\cal O}_2(t_2) \dots \rangle = Z^{-1} \mbox{Tr}\,
T[{\cal O}_1(t_1) {\cal O}_2(t_2) e^{-\beta H}]  \; ,
\label{eq5}
\end{equation}
where $\beta$ can be interpreted as an extent in Euclidean time,
$T[\dots]$ stands for time ordering of the operators inside the square
bracket with respect to the Euclidean evolution implemented by 
$\exp(-\beta H)$, and $Z=\mbox{Tr}\, \exp(-\beta H)$ is the partition
function.

\section{Path integral form}
Our goal is to provide an equivalent path integral formulation of
Eq.~\ref{eq5} conducive to calculation by numerical simulation,
following a rather standard procedure to convert from the Hamiltonian
into a Lagrangian.  We will first express the expectation values and
the partition function in terms of an integral over anticommuting
fermionic fields, i.e.~elements of a Grassmann algebra.  (The
literature on the path integral formulation of quantum expectation
values is very rich.  In our work we followed the very clear and
useful formulation given in the first chapters
of~\cite{Negele-Orland}.)  This gives origin to an integrand with an
exponential containing a quadratic form in the fermionic fields, from
$H_2$ and the normal ordering of $H_C$, as well as a quartic
expression, from $H_C$.  The quartic expression can be reduced to a
quadratic form by a Hubbard-Stratonovich transformation~\cite{HST},
through the introduction of a suitable auxiliary bosonic field (in our
case a real field), and now the Gaussian integral over the elements of
the fermionic variables can be explicitly performed, leaving an
integral over the bosonic field only.  The problem, however, is to
obtain an integral that can be interpreted as an integration over a
well defined probabilistic measure, which can thus be approximated by
stochastic simulation techniques.  We will show here how the
symmetries of the system make this possible.

We start by rewriting the expression for the charge as
\begin{equation}
q_x=a^\dag_{x,1} a_{x,1}-a_{x,-1} a^\dag_{x,-1}.
\label{eq6}
\end{equation}
We now introduce hole creation and annihilation operators for the
electrons with spin $-1$:
\begin{equation}
b^\dag_x = a_{x, -1}, \quad b_x=a^\dag_{x,-1}
\label{eq7}
\end{equation}
so that the charge becomes
\begin{equation}
q_x=a^\dag_x a_x-b^\dag_x b_x.
\label{eq8}
\end{equation}
Note that we dropped the spin indices since from now on $a, a^\dag$ and
$b, b^\dag$ will always refer to spin 1 and $-1$, respectively.
Finally we change the sign of the $b, b^\dag$ operators on one of the
sublattices.  The crucial constraint is that all redefinitions of the
operators respect the anticommutator algebra.
From the fact that $H_2$ only couples sites on the
two different sublattices, it follows that $H_2$ now takes the form
\begin{equation}
H_2=   \sum_{\langle x,y\rangle} -\kappa (a^\dag_x a_y + b^\dag_x b_y + \mbox{h.c.}).
\label{eq9}
\end{equation}

We introduce fermionic coherent states
\begin{eqnarray}
\vert \psi, \eta \rangle &=& 
e^{-\sum_x (\psi_x a^\dag_x+\eta_x b^\dag_x)} \vert 0 \rangle, 
\nonumber \\
\langle \psi^*, \eta^* \vert &=& \langle 0 \vert
e^{-\sum_x (a_x  \psi_x + b_x \eta_x)}
\label{eq10}
\end{eqnarray}
where $\psi_x, \psi_x^*, \eta_x, \eta_x^*$ are anticommuting
fermionic variables (elements of a Grassmann algebra).

The path integral formulation is obtained by factoring
\begin{equation}
e^{-\beta H} =  e^{- H\, \delta} e^{- H\, \delta} \dots e^{- H\, \delta}
\quad (N_t\; \mbox{terms})
\label{eq11}
\end{equation}
with $\delta = \beta /N_t$, and then inserting repeatedly among
the factors the resolution of the identity expressed in terms 
of an integral over the fermionic variables. The trace
in Eq.~\ref{eq5} must also be expressed in terms of a similar
integral. (See e.g.~\cite{Negele-Orland} for details.)  This leads to
integrals over fermionic fields $\psi_{x,t}, \psi_{x,t}^*, 
\eta_{x,t}, \eta_{x,t}^*$ (the index $t=0,N_t-1$ appears because of the
multiple resolutions of the identity and can be thought of as an
index labeling Euclidean time), which contain in the integrand
expressions of the type
\begin{equation}
\langle \psi^*_{x,t}, \eta^*_{x,t} \vert e^{-H \,\delta} \vert
\psi_{x,t}, \eta_{x,t} \rangle.
\label{eq12}
\end{equation}
The last ingredient is the identity
\begin{equation}
  \begin{split}
    & \langle \psi^*_{x,t}, \eta^*_{x,t} \vert F(a_x^\dag,b_x^\dag,a_x,b_x) \vert \psi_{x,t}, \eta_{x,t} \rangle \\
    & \quad = F(\psi_{x,t}^*, \eta_{x,t}^*, \psi_{x,t}, \eta_{x,t})e^{\sum_x(\psi^*_{x,t} \psi_{x,t} + \eta^*_{x,t}   \eta_{x,t})}
  \end{split}
\label{eq13}
\end{equation}
which is true of any normal ordered function $F$ of the operators
$a_x^\dag,b_x^\dag, a_x,b_x$. 

The Hamiltonian is in fact already in
normal order except for the local term $e^2 V_{xx} q_x q_x$, which
can be written as the sum of two normal-ordered pieces,
\begin{equation}
e^2 V_{xx} q_x q_x = e^2 V_{xx} :q_x q_x: + \; 
e^2 V_{xx} (a^\dag_x a_x + b^\dag_x b_x).
\label{eqNormal}
\end{equation}
By reassigning the quadratic term in Eq.~\ref{eqNormal} to $H_2$, the
exponent $-H \,\delta$ in Eq.~\ref{eq12} is normal ordered but the exponential
$\exp(-H \,\delta)$ is not.  However $\exp(-H \,\delta)$ differs
from its normal ordered form $:\exp(-H \,\delta):$ by terms $O(\delta^2)$.
So, in the limit of $N_t \to \infty$ one may replace the operator
expression $\exp(-H \,\delta)$ with an exponential involving the
fermionic fields, as follows from Eq.~\ref{eq13}. This leads to
the following expression for the partition function 
\begin{align}
\label{eq14}
  Z & = \lim_{N_t\to \infty} \int \prod_{m=0}^{N_t-1} d\psi_m^* d\psi_m d\eta_m^* d\eta_m \\
  & \!\times e^{-\sum_{m,n}( \psi^*_m M_{m,n} \psi_n+\eta^*_m M_{m,n} \eta_n)} e^{-\sum_{x,y,t} e^2 Q_{x,t} V_{x,y} Q_{y,t} \delta } \nonumber
\end{align}
where $Q_{x,t} =\psi^*_{x,t} \psi_{x,t} - \eta^*_{x,t} \eta_{x,t}$ and
we have used $m$ (or $n$) as a shorthand for the indices $x,t$.
$M$ is a matrix whose components may be deduced from
\begin{align}
  \label{eq15}
  \sum_{m,n} & \psi^*_m M_{m,n} \psi_n  = \sum_t\Big[\sum_x \psi^*_{x,t} (\psi_{x,t+1}-\psi_{x,t}) \\
  & + e^2 V_{xx} \psi^*_{x,t} \psi_{x,t} - \kappa \sum_{\langle x,y\rangle} (\psi^*_{x,t}\psi_{y,t}+\psi^*_{y,t}\psi_{x,t})\,\delta\Big] \nonumber
\end{align}
where $\psi_{x,N_t}$ must be identified with $-\psi_{x,0}$.

We now perform  a Hubbard-Stratonovich transformation, introducing
c-number real variables $\phi_{x,t}$ to recast the exponential
with the quartic term in the form
\begin{equation}
  \begin{split}
    & e^{-\sum_{x,y,t} e^2 Q_{x,t} V_{x,y} Q_{y,t} \delta } = \int \prod_{x,t} d\phi_{x,t} \\
    & \quad \times e^{-\sum_{x,y,t} \phi_{x,t} (V^{-1})_{x,y} \phi_{y,t} \delta/4}e^{-\sum_{x,t} \imath e \phi_{x,t}Q_{x,t}\delta },
  \end{split}
\label{eq16}
\end{equation}
where we have absorbed a constant measure factor in the definition
of the integral over $\phi_{x,t}$.

Inserting the r.h.s.~of Eq.~\ref{eq16} into Eq.~\ref{eq14} we get
\begin{align}
  Z = & \lim_{N_t\to \infty} \int \prod_{x,t} d\psi_{x,t}^* d\psi_{x,t} d\eta_{x,t}^* d\eta_{x,t} d\phi_{x,t} \nonumber \\
  & \times e^{-\sum_{x,y,t} \phi_{x,t} (V^{-1})_{x,y} \phi_{y,t} \delta/4} \nonumber \\
  & \times e^{-\sum_{x,t,y,\tau}( \psi^*_{x,t} M_{x,t;,y,\tau} \psi_{y,\tau}+\eta^*_{x,t} M_{x,t;y,\tau} \eta_{y,\tau})} \nonumber \\
  & \times e^{-\sum_{x,t} \imath e \phi_{x,t} (\psi^*_{x,t} \psi_{x,t} -\eta^*_{x,t} \eta_{x,t}) \delta }.
\label{eq17}
\end{align}
It is convenient to introduce a matrix $\Phi$ which is diagonal,
with diagonal entries
\begin{equation}
\Phi_{x,t} = \phi_{x,t}\delta.
\label{eq18}
\end{equation}
With this, Eq.~\ref{eq17} can be written in very compact form
\begin{equation}
  \begin{split}
    Z = \int & d\phi d\psi^* d\psi d\eta^* d\eta \\
    & \times e^{- \phi V^{-1} \phi  \delta/4 - \psi^* (M+\imath e \Phi) \psi -\eta^* (M-\imath e \Phi) \eta}
  \end{split}
\label{eq19}
\end{equation}
where we have used matrix notation for all the sums and have dropped
the limit notation.

The Gaussian integration over the anticommuting variables can now
be done to obtain
\begin{equation}
Z=  \int  d\phi e^{- \phi V^{-1} \phi  \delta/4}
{\rm det}(M-\imath e \Phi)
{\rm det}(M+\imath e \Phi).
\label{eq20}
\end{equation}
Because of the identity,
\begin{equation*}
  \det(M-\imath e \Phi)\det(M+\imath e \Phi) = \det[(M+\imath e \Phi)^\dag (M+\imath e \Phi)]
\end{equation*}
the measure is positive definite. The down spins are treated
as antiparticles (holes) moving backward in time relative to 
the up spins, exactly canceling the phase for each separately.
Correlators for the fermion operators are now be obtained by 
integrating the appropriate matrix elements of $(M+\imath e \Phi)^{-1}$
or $(M-\imath e \Phi)^{-1}$ with the measure given by Eq.~\ref{eq20}.

Equation~\ref{eq20} is the main result of our work.  It establishes
the partition function and expectation values as integrals over
real variables with a positive definite measure.  This is a crucial step
for the application of stochastic approximation methods.  There
remains the problem of sampling the field $\phi_{x,t}$ with a measure
which contains the determinant of a large matrix.  But, following
what is done in lattice gauge theory, this challenge
can be overcome through the application of the hybrid
Monte Carlo (HMC) technique~\cite{HMC}.  In a broad outline, in HMC
one first replaces the determinants in Eq.~\ref{eq20} with a Gaussian 
integral over complex pseudofermionic variables $\zeta_{x,t}$:
\begin{equation}
  \begin{split}
  \det & \big[(M+\imath e \Phi)^\dag (M+\imath e \Phi)\big] \\
  & = \int  d\zeta^* d\zeta e^{- \zeta^* (M+\imath e \Phi)^{\dag \, -1 } (M+\imath e \Phi)^{-1} \zeta}.
  \end{split}
  \label{eq21}
\end{equation}
(In this equation and in the following Eq.~\ref{eq22} we absorb 
an irrelevant, constant measure factor in the definition of the integrals.)
One then introduces real ``momentum variables'' $\pi_{x,t}$
conjugate to $\phi_{x,t}$ and inserts in Eq.~\ref{eq21} unity written
as a Gaussian integral over $\pi$.  One finally arrives at
\begin{equation}
  \begin{split}
    Z = & \int  d\phi d\pi d\zeta^* d\zeta \\
    & \times e^{- \phi V^{-1} \phi  \delta/4 - \zeta^* (M+\imath e \Phi)^{\dag \, -1} (M+\imath e \Phi)^{-1} \zeta - \pi^2/2}.
  \end{split}
  \label{eq22}
\end{equation}
The idea of HMC  is to consider the simultaneous distribution
of the variables $\phi, \pi, \zeta$ and $\zeta^*$ determined by the
measure in Eq.~\ref{eq22}.  The phase space of these variables
is explored by first extracting the $\pi$, $\zeta$ and $\zeta^*$ 
according to their Gaussian measure, and then evolving the $\phi$
and $\pi$ variables with fixed $\zeta, \zeta^*$ according to 
the evolution determined by the Hamiltonian
\begin{equation}
{\cal H}(\pi,\phi)=\frac{\pi^2}{2}   + \frac{\phi V^{-1} \phi  \delta}{4} +
\zeta^* (M+\imath e \Phi)^{ \dag \, -1 } (M+\imath e \Phi)^{-1} \zeta.
\label{eq23}
\end{equation}
Because of Liouville's theorem, the combined motion through phase
space will produce an ensemble of variables distributed according
to the measure in Eq.~\ref{eq22} and, in particular, of fields $\phi$ 
distributed according to the measure of Eq.~\ref{eq20}.

Of course, the discussion above assumes that the Hamiltonian evolution of
$\phi$ and $\pi$ is exact, which will not be the case with a numerical
evolution.  The HMC algorithm addresses this shortcoming
by: 1) approximating the evolution with a symplectic integrator which
is reversible and preserves phase space, 2) performing a Metropolis 
accept-reject
step at the end of the evolution, based on the variation of the value
of the Hamiltonian.

\section{Numerical Tests}

We tested our method on the two-site system
obtained by taking $L=1$, which can be solved exactly.  We label
the sites $x=0, 1$. With $\kappa =1/3$, the Hamiltonian
$H = H_2 + H_C$ is now
\begin{align}
  H_2 & = -(a_1^\dag a_0 + a_0^\dag a_1 + b_1^\dag b_0 + b_0^\dag b_1) \nonumber  \\
  & \quad + \mu (a^\dag_0 a_0 + b^\dag_0 b_0 + a^\dag_1 a_1 + b^\dag_1 b_1) \\
  H_C & = 2 e^2  (a_0^\dag a_0 - b_0^\dag b_0) (a_1^\dag a_1 -
  b_1^\dag b_1) +  \frac{2  e^2}{r_0} a_0^\dag b_0^\dag a_0  b_0  \; , \nonumber
\label{eq24}
\end{align}
where we have taken $V_{0,1}=V_{1,0}=1/3$ and a local interaction term
$V_{0,0}=V_{1,1}=1/r_0$.  The radius $r_0$ sets the physical scale
in lattice units for localization of the net charge at the carbon atom.
It must be restricted to $r_0 < 1$ for stability of the vacuum. Also the normal
ordering prescription for $e^2 V_{xx}q_x q_x$ in Eq.~\ref{eqNormal}
adds a new contribution to $H_2$ in the form of an $I_3$ ``chemical
potential'' $\mu a^\dag_{x,s} \sigma^{ss'}_3 a_{x,s'}$.  It is well
known~\cite{iso} that an $I_3$ chemical potential for any value of
$\mu$ does not introduce a phase in the measure.  To maintain the
full $SU(2)$ ``flavor'' symmetry of the tight-binding graphene
Hamiltonian, we must set $\mu = e^2/r_0$.  For the two-site
system, the isospin generators of Eq.~\ref{eq:iso} become
\begin{equation*}
  I_+ = I^\dag_- = (-1)^x a^\dag_x b^\dag_x \mbox{ and } I_3 = [a^\dag_x a_x + b^\dag_x b_x]/2 -1,
\end{equation*}
allowing us to unambiguously classify the 16 states as
degenerate $I =0,1/2,1$ isomultiplets: 5 singlets, 4
doublets and one triplet.

\begin{figure}[t]
  \centering
  \includegraphics[width=8cm]{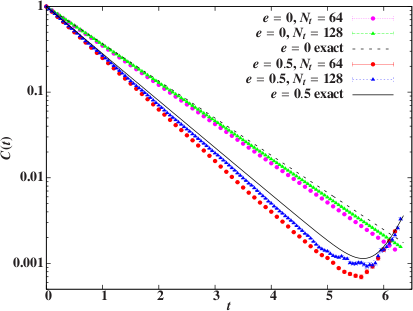}
  \caption{HMC data with $N_t = 64$ and 128 converging to exact results for the
    correlation function of Eq.~\ref{eq25a} with $\mu = e^2/r_0$, $r_0 = 1/2$ and
    fixed $\beta = N_t \delta = 6.4$.  The upper set of results are for $e = 0$,
    the lower set for $e = 0.5$.}
  \label{fig2}
\end{figure}

We compared HMC results for expectation values of several
products of fermionic operators with the corresponding
exact values, finding satisfactory agreement.
For example, the correlation function
\begin{equation}
C_a(t)=\langle (a_0-a_1)(t) \, (a_0^{\dag}-a_1^{\dag})(0)\rangle /2
\label{eq25a}
\end{equation}
is illustrated in Fig.~\ref{fig2}, which shows HMC results converging
to the exact correlators for both the free theory with $e = 0$
and an interacting case with $e = 0.5$.

\begin{figure}[t]
  \centering
  \includegraphics[width=8cm]{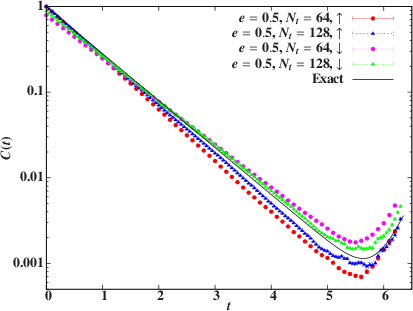}
  \caption{HMC data with $N_t = 64$ and $N_t = 128$ for the correlation
  functions of Eqs.~\ref{eq25a} (``$\uparrow$'', red squares and blue
  triangles) and \ref{eq25b} (``$\downarrow$'', pink squares and green
  triangles) converging to the same exact continuum result (black solid line).
  Simulation parameters are the same as in Fig.~\ref{fig2}.}
  \label{fig3}
\end{figure}

A stringent test is to demonstrate the convergence to exact $SU(2)$
symmetry in the ``time'' continuum limit.  To this end, consider a
second correlation function,
\begin{equation}
  C_b(t)=\langle (b_0^{\dag} + b_1^{\dag})(t) \, (b_0 + b_1)(0)\rangle /2 \; ,
\label{eq25b}
\end{equation}
related to $C_a(t)$ by an $SU(2)$ rotation. In Fig.~\ref{fig3} we
compare HMC and exact results for both $C_a(t)$ and $C_b(t)$, finding that
both correlators converge to the same continuum result.

\begin{figure}[t]
  \centering
  \includegraphics[width=8cm]{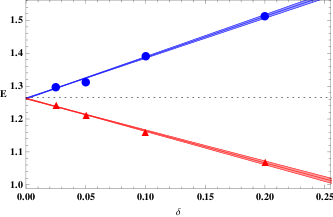}
  \caption{Linear extrapolation (including error band) of the HMC energies for
    isodoublet correlators $C_a(t)$ (blue dots) and $C_b$ (red triangles) as a function of
    the ``time'' lattice spacing $\delta = 6.4/N_t$ for $N_t = 32$,
    $64$, $128$ and $256$.
    The dotted horizontal line marks the exact continuum result.}
  \label{fig4}
\end{figure}

We extract the energies of the isodoublet states at nonzero
$\delta = \beta/N_t$ by fitting the correlator data in
Fig~\ref{fig3} to single exponentials, $C(t) \approx e^{-Et}$ for
fit range $0.4 < t < 4$. The results in Fig.~\ref{fig4} clearly show
linear behavior $E \approx E_0 + c_1\delta$, converging to the exact continuum
energy is $E_0 = e^2 + \sqrt{4 + e^4} - 1 \approx 1.266$.
The continuum limit is consistent with restoration of the  $SU(2)$
symmetry of the Hamiltonian: a simultaneous linear fit to both sets of energies
gives $\lim_{\delta \to 0} E = 1.262 \pm 0.004$, with $c_1 = 1.25 \pm 0.07$
and $-0.98 \pm 0.04$ for correlators $C_a$ and $C_b$, respectively.

\section{Conclusion}

While the results we reported are for a small test system, they demonstrate
that HMC simulations of graphene directly
on the hexagonal graphene lattice are possible and have the potential to
produce valuable results.  The dominant nearest neighbor hopping term
has no sign problem, and we anticipate that
a small next-to-nearest neighbor coupling $\kappa'/\kappa \simeq 0.05$
can be accommodated by reweighting without a prohibitive cost. The crucial
observation is the cancellation between the phase of the up spin and
down spin determinant, when the latter are treated as holes moving
backward in time. We are currently pursuing simulations
of larger systems, and beginning to explore the many possible
investigations and generalizations
(e.g., distortions of the lattice, phonons, inclusion of magnetic fields).

{\bf Acknowledgments:} We wish to acknowledge the many fruitful
conversations with Dr. Ronald Babich, Prof. Antonio Castro Neto and
Prof. Claudio Chamon during the course of this research and support
under DOE grants DE-FG02-91ER40676,
DE-FC02-06ER41440, and NSF grants OCI-0749317, OCI-0749202. Part of
this work was completed while two of the authors were at the Aspen
Center for Physics.

\end{document}